\documentclass[aps,prb,reprint,superscriptaddress]{revtex4-2}
\usepackage{graphicx}  % need for figures
\usepackage{xcolor} % for editing only
\usepackage{amsmath}
\usepackage{tabularx}
\usepackage{subfigure}
\usepackage{float}
\usepackage{bm} % bold math
\begin{document}
	\title{Determination of the Magnetic Penetration Depth from Finite-Element Susceptibility Calculations of Spheroidal and Cylindrical Superconductors}
	\author{M. J. Grant}
	\author{A. Carrington}
	\affiliation{H. H. Wills Physics Laboratory, University of Bristol, Tyndall Avenue, Bristol, BS8 1TL, UK.}

\date{\today}

\begin{abstract}

The magnetic penetration depth, $\lambda$, is a fundamental parameter of the superconducting state and is commonly determined from precision measurements of the magnetic susceptibility, $\chi$, in the Meissner state. However, extracting $\lambda$ from such measurements requires an accurate treatment of geometry-dependent demagnetization effects, for which analytic results are only available in a limited number of cases. Here we present a finite-element analysis of the magnetic response of three-dimensional superconducting spheroids and cylinders with finite penetration depth. By calculating the magnetic moment and susceptibility, $\chi(\lambda)$, over a wide range of aspect ratios, we quantify the shape-dependent Meissner response and extract effective demagnetization factors in the $\lambda \rightarrow 0$ limit. For small but finite $\lambda$, we further determine effective sample dimensions from the initial variation of $\chi(\lambda)$ and propose an improved analytical expression for these. Our results provide a high-precision procedure for determining $\lambda$ from volume-exclusion based susceptibility measurements and should be directly applicable to the characterization of novel superconducting materials.

\end{abstract}

\maketitle
	
\section{Introduction}
 
In the Meissner state of a superconductor, geometry-dependent demagnetisation effects distort the magnetic field both at the surface of the sample and in the surrounding region. While these effects can locally enhance or reduce the magnetic field, the overall effect is an enhancement of the magnitude of the volume susceptibility, $\chi$, such that in the limit that the magnetic penetration depth, $\lambda=0$, $\chi<-1$. A finite value of $\lambda$ reduces $|\chi|$ and this reduction is the basis of common methods for the experimental determination of $\lambda$, particularly its variation with temperature \cite{giannetta22}. 

For ellipsoidal geometries in a uniform applied field, the internal demagnetizing field is uniform, and a constant scalar demagnetization factor, $N$, relates it to the magnetization \cite{osborn45,landau84}. For non-ellipsoidal geometries, however, both the magnetization and the demagnetizing field are non-uniform and so no unique scalar demagnetization factor exists. Instead, effective demagnetization factors can be defined for specific applications, such as calculating the total magnetic moment. Such calculations have been performed for various geometries, including cylinders, cuboids, slabs, and rectangular prisms \cite{joseph65,chen91,aharoni98,pardo04,sato89,smith10}. However, these treatments often rely on simplifying assumptions about the magnetization, and their applicability to superconductors is therefore uncertain \cite{prozorov18}. More recently, Prozorov and Kogan \cite{prozorov18} have used finite element analysis to solve Maxwell's equations in three dimensions for a large range of geometries to determine superconducting effective demagnetization factors in the limit $\lambda=0$.

To determine $\lambda$ from the measured value of $\chi$ for a particular sample geometry requires accurate knowledge of the functional form of $\chi(\lambda)$.  For  a few special cases, there are analytical solutions to the London equation for finite $\lambda$. These include an infinite slab of width $2a$ (1D), an infinite cylinder of radius $a$ (2D) and a sphere of radius $a$ (3D) \cite{london61}.  In each of these cases, it is found that for $\lambda\ll a$, 
\begin{equation}\label{EqChiLinear} 
   \chi(\lambda) = -\frac{1}{1-N} \left( 1 - \frac{\lambda}{R}\right),
\end{equation}
where $R$ is an effective sample dimension.  However, for typical sample geometries, which are often thin flat parallelepipeds, these analytical cases offer only a very approximate solution.  

A marked improvement on these simple approximations was made by Prozorov \textit{et al}. \cite{prozorov00}, who recognized that although an accurate calculation of $N$ was difficult for typical flat plate-like samples with $N\simeq 1$, this difficulty could be reduced by a simple experimental procedure. For a sample of volume, $V_s$, when $\lambda$ is much less than all sample dimensions, $V_s/(1-N)$ can be estimated by removing the sample from the measurement coil or cavity. Changes in $\chi$, for example, as a function of temperature, can then be directly related to changes in $\lambda$ provided $R$ [Eq.\ (\ref{EqChiLinear})] is known. Estimates based on two-dimensional numerical calculations, together with intuition gained from the known analytic solutions, suggested that $R$ varies only weakly with the aspect ratio of the sample, and an approximate variation of $R$ as a function of the aspect ratio was derived. This estimated solution for $R$ has been widely used in analyzing experimental measurements of $\lambda$.  More recently, Prozorov \cite{prozorov21} has presented a comprehensive finite element analysis of demagnetization effects in cylindrical superconductors, accurately determining $\chi(\lambda/a)$ and estimating $R$ for this geometry.

In the present work, we report our finite element analysis of demagnetization effects in superconducting spheroids and cylinders. By solving Maxwell's equations and the London equation in three dimensions, we compute the full spatial dependence of the magnetic vector potential and extract $\chi(\lambda)$ with high numerical accuracy. This enables us to extract $\chi(\lambda)$, the effective demagnetization factor, $N$, and the effective dimension, $R$ for these geometries. Our determination of $R$ has improved accuracy over previous works\ \cite{prozorov21} and thus improves the accuracy of the extraction of $\lambda$ from experimental data. 

\section{Methods}

For linear magnetic materials, the magnetization, $\bm{M}$, is linearly related to the magnetic field, $\bm{H}$, via the intrinsic susceptibility, $\chi_0$, $\bm{M}=\chi_0 \bm{H}$. The magnetic induction, $\bm{B}$ is then given by ${\bm{B}=\mu_0(\bm{H}+\bm{M})}$. In the presence of an applied field $\bm{H}_0$, the induced magnetization produces a demagnetizing field, $\bm{H}_{\mathrm{D}}$ \cite{jackson}. In the special case of an ellipsoid, $\bm{M}$ and $\bm{H}_{\mathrm{D}}$ are spatially uniform and related by a constant demagnetization factor $0\leq N\leq 1$ \cite{osborn45,jackson}. The demagnetisation factors for field aligned along each principal axis satisfy a sum rule; $N_a + N_b + N_c =1$. The internal field of an ellipsoid is then $\bm{H}_{\mathrm{int}} = \bm{H}_0 + \bm{H}_{\mathrm{D}} = \bm{H}_0 - N\bm{M}$ and an effective susceptibility can be defined,
\begin{equation}\label{EqDefDemag}
	\bm{M} = \frac{\chi_0}{1+\chi_0 N}\bm{H}_0 = \chi \bm{H}_0\ ,
\end{equation}
such that demagnetization effects are incorporated in $\chi$.

In typical single crystal samples of superconductors, which are not ellipsoids, the magnetization is non-uniform and there is no single constant $N$. Nevertheless, a volume average demagnetization factor, which relates $\chi$ and $\chi_0$, can be defined by insisting that Eq.\ (\ref{EqDefDemag}) holds when $\bm{M}$ is replaced by its volume-averaged value \cite{prozorov18}. In this generalized context the sum-rule no longer applies.

More formally, the total magnetic moment of an arbitrary current distribution, $\bm{J}(\bm{r})$, is
\begin{equation}\label{EqmfromJ}
	\bm{m} = \frac{1}{2} \int \bm{r}\times \bm{J}(\bm{r})\, dV_{\mathrm{s}} \ ,
\end{equation} 
where the integral is performed over the sample volume, $V_{\mathrm{s}}$ \cite{jackson}. In a superconductor the screening currents obey the London relation $\bm{J}=-\bm{A}/\mu_0\lambda^2$ \cite{london61}, so that in cylindrical coordinates, for an applied field $\bm{H}_0 = H_0 \hat{\bm{z}}$, the susceptibility is
\begin{equation}\label{EqchiA}
	\chi = -\frac{1}{2\mu_0{{H}_0}\lambda^2 V_{\mathrm{s}}}\int rA_{\phi}\, dV_{\mathrm{s}}\ .
\end{equation}
Finally, to be consistent with the conventional definition of the demagnetization factor, we define $N$ in the $\lambda\rightarrow 0$ limit where $\chi_0 = -1$,
\begin{equation}\label{EqNdef}
	N = 1-\frac{1}{\chi(\lambda=0)} \, .
\end{equation}

\begin{figure}
	\centering
	\includegraphics[width=0.8\columnwidth]{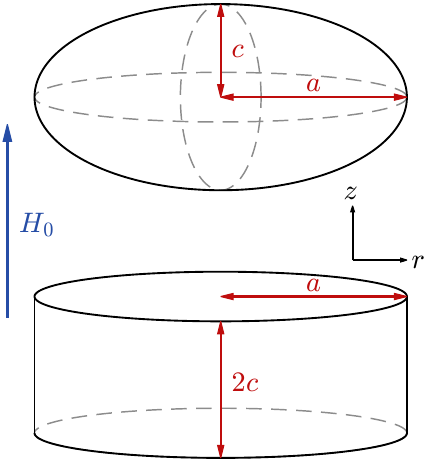}
	\caption{Geometry of superconducting samples: (top) Spheroid ($a\times a \times c$), (bottom) finite cylinder (radius $a$, length $2c$). The applied field is aligned along the $c$-axis.  The aspect ratio $\nu=c/a$.}
	\label{figGeometry}
\end{figure}

In our study we focus on two superconducting geometries; a spheroid with semi-axes $a\times a\times c$, and a finite cylinder (`disk') of radius $a$ and thickness $2c$. Their orientations, with field always applied along the $c$-axis (the axis of revolution), are illustrated in Fig. \ref{figGeometry}. For the remainder of the article we will refer to the aspect ratio, $\nu=c/a$.

The calculations were performed using \textit{COMSOL Multiphysics} \cite{comsol}; specifically the \textit{AC/DC module} to solve for the magnetic vector potential, $\bm{A}(\bm{r})$. This solves the frequency-dependent, $\omega$, form of Amp\`{e}re's law,
\begin{equation}\label{EqComsolPde} 
	\bm{\nabla}\times(\bm{\nabla}\times\bm{A}) =
	 -i\omega\sigma\mu_r\mu_0\bm{A} + \omega^2\varepsilon_r\varepsilon_0\mu_r\mu_0\bm{A} \ ,
\end{equation} 
where $\sigma$ is the electrical conductivity. By using a complex, frequency-dependent conductivity of the form
\begin{equation} \label{EqSigmaSubs}
		\sigma = -\frac{i}{\mu_0\omega\lambda^2} 
\end{equation}
and setting $\mu_r=1\ $, $\varepsilon_r=0$, we transform Eq.\ (\ref{EqComsolPde}) into the London equation $\bm{\nabla}\times(\bm{\nabla}\times\bm{A}) = -\bm{A}/\lambda^{2}$ and use this to model the interior of the superconductor. The region outside of the superconductor is modeled as a vacuum ($\sigma=0$, $\varepsilon_r = \mu_r = 1$). In our simulations, the samples are axially symmetric which significantly reduces the computational resources required.  

A uniform applied field along $\hat{\bm{z}}$ is introduced using \textit{COMSOL}'s reduced-field formulation, with background vector potential $\bm{A}=\mu_0(0,H_0r/2,0)$. This condition is imposed on an outer spherical boundary of radius $t\gg a$, chosen to avoid truncating the field enhancement in the vicinity of the sample. To quantify the effect of the finite outer boundary, convergence tests were performed by increasing $t$ up to $50a$ for the limiting values of $\nu$ and $\lambda/a$ considered here. Increasing $t$ from $5a$ to $50a$ changed the derived susceptibility by at most $0.45\%$ for spheroids and $0.75\%$ for disks. For the results presented here, we used $t=30a$ and estimate the boundary-induced uncertainty in $\chi$ to be less than $0.001\%$.

\begin{figure}
	\centering
	\includegraphics[width=\columnwidth]{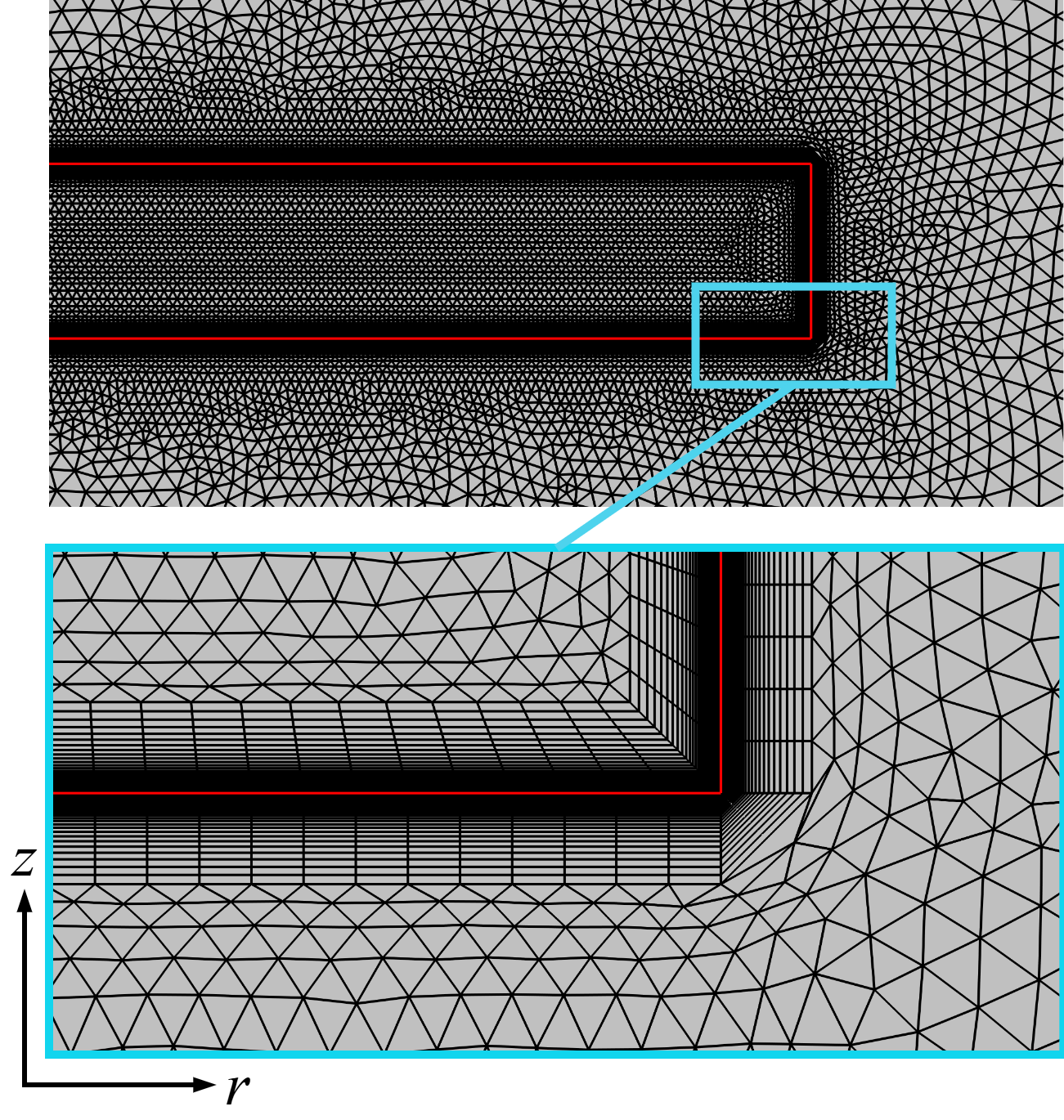}
	\caption{Axisymmetric mesh for the cylindrical geometry. The majority of the model is meshed using triangular elements; the inset shows the dense boundary-layer mesh at the superconductor surface.}
	\label{figMesh}
\end{figure}

Accurately resolving the demagnetization-enhanced fields outside the sample, together with the exponential decay of $\bm{A}$ within a distance $\sim\lambda$ of the surface, requires substantial mesh refinement near the superconductor--vacuum interface. The bulk of the superconducting computational domain was meshed using triangular elements with maximum size $0.01a$. At the interface, a quadrilateral boundary-layer mesh was used, with first-layer thickness $\lambda/100$, a growth rate of $\sim10\%$ in successive layers, and a total thickness of approximately $3\lambda$. A schematic of the mesh for the disk geometry is shown in Fig.~\ref{figMesh}. Systematic variation of the mesh size, boundary-layer thickness, and additional corner refinement confirmed convergence of the calculated susceptibility.

\begin{figure}
	\centering
	\includegraphics[width=\columnwidth]{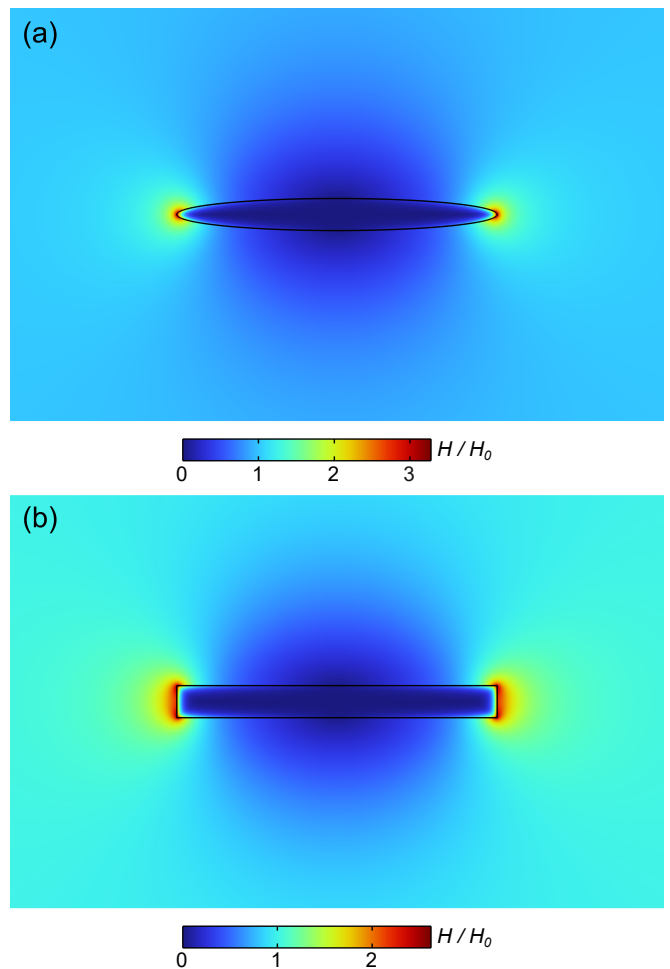}
	\caption{Calculated magnitude of the magnetic field, $H/H_0$, for (a) spheroid and (b) cylinder both with $\nu=c/a=0.1$ and $\lambda/a=0.025$.}
	\label{figMagneticField}
\end{figure}

\begin{figure}
	\centering

		\includegraphics[width=\columnwidth]{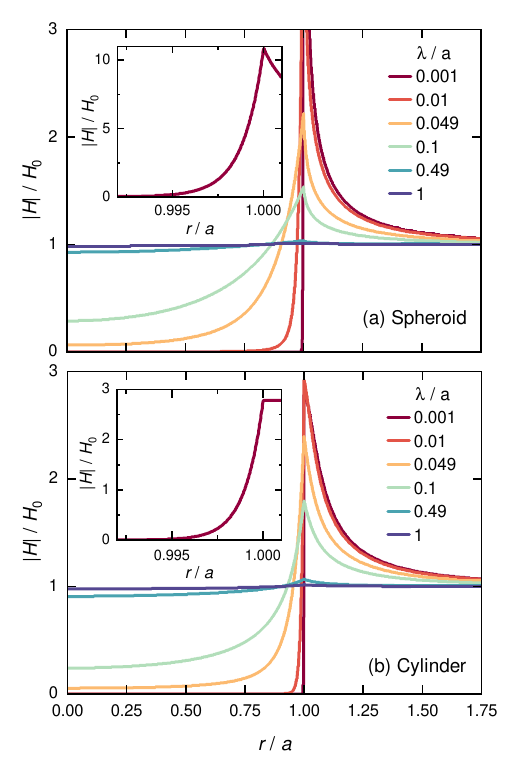}
	\caption{Radial field profiles at $z=0$ for (a) spheroid and (b) cylinder with $c/a=0.05$ for various $\lambda/a$. The insets highlight the $\lambda=0.001a$ case near the surface.}
	\label{figRadialField} 
\end{figure}

Fig.\ \ref{figMagneticField} shows example maps of the magnitude of the magnetic field in the $r-z$ plane for $\nu=0.1$ and $\lambda/a=0.025$; showing where the field is reduced in the interior of the sample and the regions above the axial line (dark blue), and where the field is enhanced near the extremity of the sample along the radial axis (dark red). Fig. \ref{figRadialField} shows radial field profiles at $z=0$ for $c/a=0.05$ and several values of $\lambda/a$. The insets to these figures show the profiles near the surface for the case where $\lambda/a=0.001$. The smooth evolution of the field demonstrates that, even at the smallest $\lambda /a$ values considered in our simulations, the mesh density is adequate to capture the relevant features of the solution.

\section{Results}
Using the parameters outlined above, we have numerically determined the spatial distribution of $\bm{A}$, for both cylindrical and spheroidal geometries over the ranges $\nu=c/a = 0.05-3$ and $\lambda/a = 0.001-1$. From these results, we calculate the effective susceptibility, $\chi$, using Eq.\ (\ref{EqchiA}). Our method differs from that of Refs.\ \cite{prozorov18,prozorov21}, in that we determine $\bm{m}$ by integrating $\bm{A}(\bm{r})$ inside the sample as opposed to integrating $\bm{B}(\bm{r})$ over all space. We have performed a detailed comparison of $\chi(\lambda)$ between our results and that of Ref.~\cite{prozorov21} and find them to be in very good agreement, so both techniques give equivalent results provided they are well converged.

\begin{figure}
	\centering
		\includegraphics[width=\columnwidth]{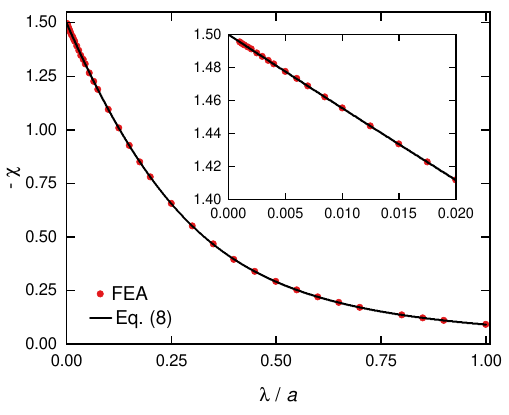}
	\caption{Numerically calculated effective susceptibility, $\chi$, of a superconducting sphere as a function of $\lambda/a$, compared with the exact analytical expression [Eq.\ (\ref{EqChiSphere})]. The inset shows the linear behavior for small $\lambda/a$.}
	\label{figChiSphere}
\end{figure}

We begin our analysis by focusing on the simple case of a sphere of radius $a$, for which an exact analytical expression for $\chi(\lambda)$ is known;
\begin{equation}\label{EqChiSphere}
		\chi = -\frac{3}{2}\left( 1 - \frac{3\lambda}{a}\coth\left(\frac{a}{\lambda}\right)+3\left(\frac{\lambda}{a}\right)^2\right)\ .
\end{equation}
Here $\chi = -3/2$ in the $\lambda=0$ limit, which corresponds to $N = 1/3$ \cite{london61,may59}. Our numerically calculated values are compared with this solution in Fig. \ref{figChiSphere}, showing excellent agreement. The maximum deviation from the analytical expression is approximately $\simeq0.004\,\%$, occurring at the smallest value of $\lambda/a = 0.001$.

In the $\lambda/a \ll 1$ limit, the expansion of the solution for a sphere, given by Eq.\ (\ref{EqChiSphere}), up to second order terms in $\lambda$ is
\begin{equation}  \label{EqChi2}
	\chi \simeq -\frac{1}{1-N}\left[1 - \frac{\lambda}{R} +\left(\frac{\lambda}{R_2}\right)^2\right]\ ,
\end{equation}
with $N=R/a =(R_2/a)^2 =  1/3$. Fitting our numerical values of $\chi(\lambda)$ for a sphere to Eq.\ (\ref{EqChi2}) in the range $\lambda/a = 0.001 - 0.01$, with $N$, $R$ and $R_2$ as free parameters, yields $N = 0.33336$, $R/a=0.33332$, and $(R_2/a)^2=0.33330$ in excellent agreement with the analytical values.

\begin{figure}
	\centering
	\includegraphics[width=\columnwidth]{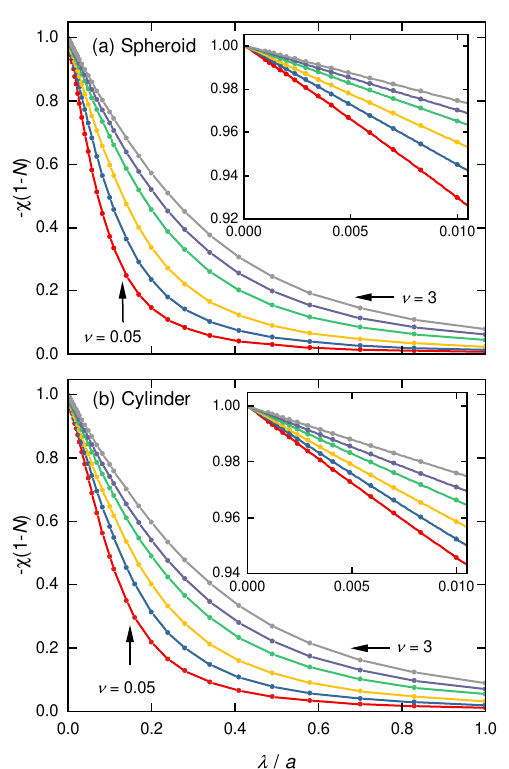}
	\caption{Examples of the scaled susceptibility, $-\chi(1 - N)$, for $\nu=0.05,\ 0.1,\ 0.2,\ 0.5,\ 1,\ 3$ and (a) spheroids and (b) cylinders. Solid lines are a guide to the eye. Insets: $-\chi(1-N)$ as a function of $\lambda/a$ in the $\lambda \ll a$ regime. Solid lines are fits to Eq.~(\ref{EqChi2}).}
	\label{figScaledChi}
\end{figure}

More generally, we find that Eq.\ (\ref{EqChi2}) provides an excellent fit to the numerically calculated susceptibility data for both spheroidal and cylindrical geometries across all aspect ratios considered up to $\lambda/a = 0.01$. This yields the effective demagnetization factor, $N$, and the effective dimension, $R$. Figure \ref{figScaledChi} shows representative examples of $\chi(\lambda)$ for both geometries, alongside fits to Eq.\ (\ref{EqChi2}) up to $\lambda=0.01a$ in the insets. 

\begin{figure}
	\centering
	\includegraphics[width=\columnwidth]{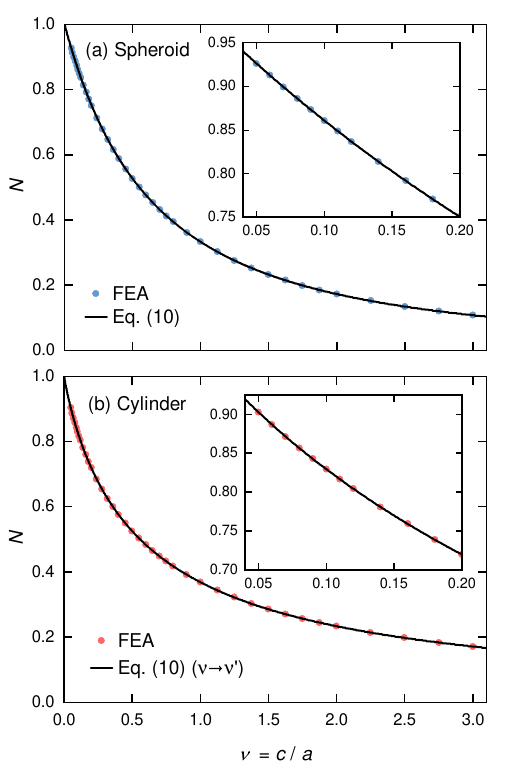}
	\caption{Demagnetization factor, $N$, extracted from numerical susceptibility data for (a) spheroids and (b) cylinders. The solid line for spheroids shows the analytical result of Eq.~(\ref{EqNSpheroid}), while for the cylinder we fit to Eq.~(\ref{EqNSpheroid}) with a rescaling of $\nu$ given by Eq.~(\ref{EqNuRescaled}). The insets are magnifications of the $\nu<0.2$ range.}
	\label{figN}
\end{figure}

The extracted values of $N$, for both spheroidal and cylindrical geometries are shown in Fig.~\ref{figN}. For the spheroidal case, we compare our finite element numerical results to the exact solution \cite{landau84},
\begin{equation}
	\label{EqNSpheroid}
	N = \frac{\nu}{2}\int_0^\infty \frac{ds}{(s+1) (s+\nu^2)^\frac{3}{2} } \ .
\end{equation}
This solution can also be expressed in an analytical form using elliptic integrals of the first and second kind \cite{osborn45}.  We find excellent agreement between the numerical results and Eq.\~(\ref{EqNSpheroid}) across the full range of aspect ratios considered; within $0.02\%$ up to $\nu=1.5$ and within $0.08\%$ up to $\nu=3$, confirming the accuracy of our methods.

For the cylindrical geometry, while no exact analytical solution for $N$ exists, our results are in excellent agreement with the calculations of Refs. \cite{prozorov18, prozorov21}. As the two calculations were performed using different methods, this gives additional confirmation of both sets of results.

\begin{figure}
	\centering
	\includegraphics[width=\columnwidth]{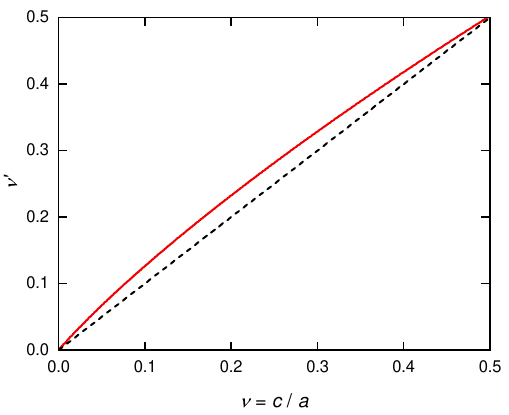}
	\caption{Form of the rescaled aspect ratio, $\nu^{\prime}$, up to $\nu=0.5$. The dashed gray line is the inscribed ellipsoid approximation; $\nu^{\prime}=\nu$.}
	\label{figRescaledAspect}
\end{figure}

As can be seen in Fig.\ \ref{figN}, the behavior of $N (\nu)$ for spheroid and cylinder is qualitatively similar. This motivates a rescaling of the aspect ratio, $\nu$, of the cylinder so that its value of $N$ is given by Eq.\ (\ref{EqNSpheroid}) with $\nu$ replaced by the rescaled ratio $\nu^\prime$. We find that in the range studied the required rescaling function is well described by
\begin{equation}
	\label{EqNuRescaled}
	\nu^\prime =  \frac{\alpha_1\nu^{\alpha_2}}{(1+\alpha_3\nu^{\alpha_4})}\ ,
\end{equation}
with coefficients
\begin{equation*}
	\begin{split}
 \alpha_1 = 2.14774\ ,\\
 \alpha_2 =  1.05029\ ,\\
 \alpha_3 =  1.45208\ ,\\
 \alpha_4 = 0.44932\ .
 	\end{split}
\end{equation*}
This parameterization reproduces the numerically determined values of $N$ to within $\pm0.01\%$ for $\nu\leq 0.5$ and within $\pm0.05\%$ up to $\nu=3$. The functional form was chosen to interpolate smoothly between the limiting behaviors at small and large $\nu$ and is shown in Fig.\ \ref{figRescaledAspect} for $\nu<0.5$.

\begin{figure}
	\centering
	\includegraphics[width=\columnwidth]{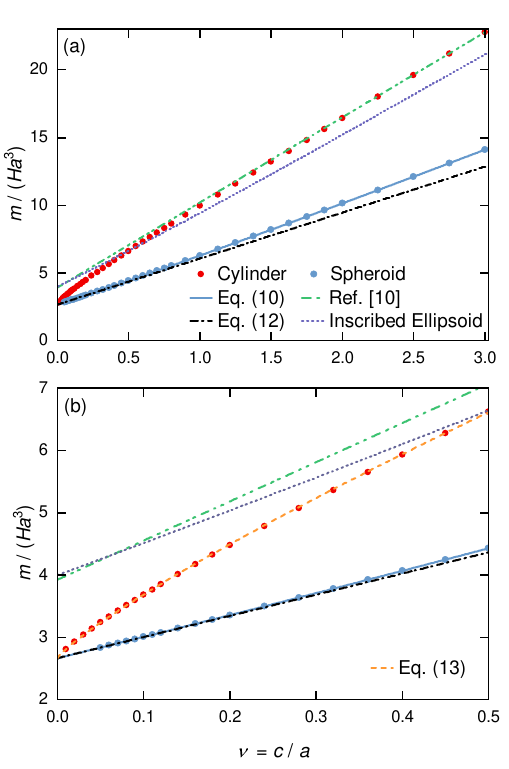}
\caption{(a) The magnetic moment per unit field,  for unit radius $a$, for both geometries.  For the spheroid, we show the FEA result alongside the exact result [Eq.\ (\ref{EqNSpheroid})] and the low $\nu$ expansion Eq.\ (\ref{EqmSpheroidLinear}). For the cylinder we show the FEA result together with the low $\nu$ approximation [Eq.\ (\ref{EqFitmcylinder})], the inscribed ellipsoid approximation and that calculated from the approximation for $N^{-1}$ suggested in Ref.\ \cite{prozorov18}. (b) Magnified view of $\nu<0.5$.}
	\label{figmagmoment}
\end{figure}

In Fig.\ \ref{figmagmoment} we show the total calculated magnetic moment per unit field in the $\lambda=0$ limit, $m/H=-V_s/(1-N)$, for both geometries. Here $V_s = \frac{4}{3}\pi a^3\nu$ for a spheroid and $V_s = 2\pi a^3\nu$ for a cylinder. In a magnetometer or ac susceptometer this is the quantity that is directly measured experimentally. For the spheroid, it follows from the expansion of Eq.\ (\ref{EqNSpheroid}) for small $\nu$, $N\simeq 1-\frac{1}{2}\pi\nu+2\nu^2$, that 
\begin{equation}\label{EqmSpheroidLinear}
    m/H\simeq \frac{8}{3}a^3\left(1+\frac{4\nu}{\pi}\right)  .
\end{equation}
A sample of aspect ratio $\nu=0.1$ has only a $13\%$ larger magnetic moment compared to that of an infinitesimally thin sample. As an infinitesimally thin spheroid and infinitesimally thin disk are indistinguishable, we believe that that for the cylinder the value of $m/H$ for $\nu=0$ should be equal to that of the spheroid and indeed this is supported by an extrapolation of the data in Fig.\ \ref{figmagmoment}. However, the rate of increase of $m/H$ for the cylinder with increasing $\nu$ is more than twice larger that of the spheroid.

A commonly used simple estimate of the magnetic moment for non-spheroidal samples is the inscribed-ellipsoid approximation. In this approach, the susceptibility enhancement factor $1/(1-N)$ is taken to be that of an ellipsoid with the same aspect ratio as the sample, and the magnetic moment is then estimated by multiplying this factor by the actual sample volume. For the cylinder, this is equivalent to assuming that $\nu^{\prime}=\nu$ in Eq.\ (\ref{EqNSpheroid}) and gives $m/H\simeq 4a^3(1+4\nu/\pi)$ for small $\nu$.   For $\nu=0$, this is $\sim$ 50\% too large, but the approximation is reasonable for $\nu>0.25$ where the deviation is less than 10\% and is approximately a constant $\sim 7\%$ for $\nu>1$. 

In Ref.\ \cite{prozorov18} a limiting form of $N$ for a cylinder with small $\nu$ was suggested to be $N\simeq (1+1.6\nu)^{-1}$, which gives $m/H=1.25\pi a^3+2\pi a^3\nu$.  This gives $m/H$ for $\nu=0$ which is almost the same as the inscribed ellipsoid approximation ($\sim 50\%$ too large),  although the agreement improves for larger $\nu$. We find a more accurate approximation to describe $m/H$ in the $\nu<0.5$ range is 
\begin{equation}\label{EqFitmcylinder}
    m/H=\frac{8}{3}a^3(1+2.65\nu^{0.84}).
\end{equation}
From this values for $N(\nu)$ for a cylinder can easily be calculated.

\begin{figure}
	\centering
	\includegraphics[width=\columnwidth]{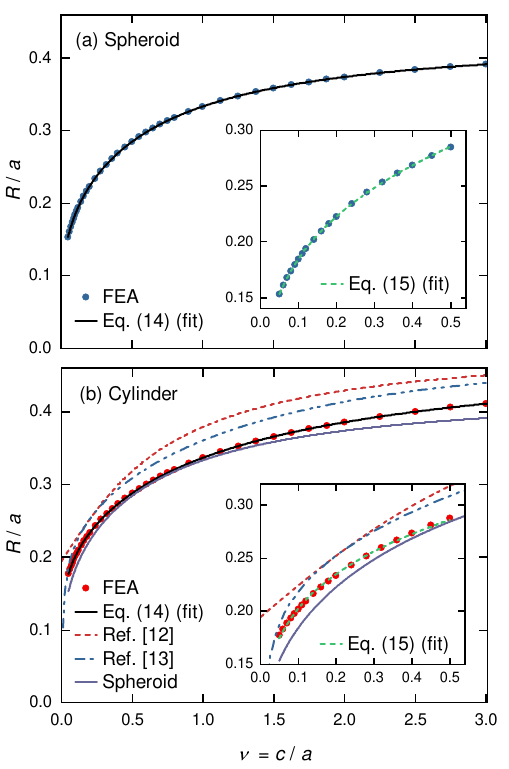}
	\caption{Effective dimension, $R$, extracted from numerical susceptibility data for (a) spheroids and (b) cylinders. The black solid lines are fits to Eq.~(\ref{EqRParam}). For cylinders we also compare to previous expressions from Refs.~\cite{prozorov00,prozorov21}.  The insets show a zoom of the low $\nu$ behavior, where we also show a fit to the simple relation Eq.\ (\ref{EqRSimpleFit}). }
	\label{figR}
\end{figure}

We now proceed to our primary focus, the effective dimension $R$, which determines the magnitude of the leading-order linear dependence of $\chi$ on $\lambda$. Fig.~\ref{figR} shows the numerically determined values of $R(\nu)$ for both spheroidal and cylindrical geometries. Unlike the case for $N$, there is no corresponding analytic solution for $R$ for a general spheroid. However, as discussed above, our results do match the analytic solution exactly in the spherical limit. In both geometries, we note that $R$ varies only weakly with $\nu$. The two cases exhibit strikingly similar behavior across the entire range, with the values for the spheroid always smaller than those for the cylinder for the same value of $\nu$. 

To accurately capture the linear-in-$\lambda$ contribution to $\chi(\lambda)$ over the full range of aspect ratios, we use the following parametrization for both geometries,
\begin{equation}
	\label{EqRParam}
 R = R_\infty\tanh{\left( \beta_1\nu^{\beta_2}+\beta_3\ln{(1+\beta_4\nu)}\right)}\ .
\end{equation}
This expression has the limit $R\rightarrow R_{\infty}$ as $\nu\rightarrow\infty$, where $R_{\infty}$ is determined by the response in the absence of demagnetization effects. $R_{\infty} =4a/3\pi$ and $R_{\infty} =a/2$ for the spheroid and cylinder respectively. The fitted parameters are, for spheroids,
\begin{equation*}
	\begin{split}
		\beta_1 =0.73233\ ,\\
		\beta_2 =0.24158\ ,\\
		\beta_3 =0.45894 \ ,\\
		\beta_4 = 1.03959\ .
	\end{split}
\end{equation*}    
and for cylinders
\begin{equation*}
	\begin{split}
		\beta_1 = 0.57196 \ ,\\
		\beta_2 = 0.16174 \ ,\\
		\beta_3 = 0.32653 \ ,\\
		\beta_4 = 1.11931 \ .
	\end{split}
\end{equation*}
This provides excellent agreement with the numerical data, with relative errors less than $\pm0.2\%$ in the entire range of $\nu$ studied. 

For $0.05<\nu<0.5$, a simpler fit is possible
\begin{equation}
\label{EqRSimpleFit}
     R=R_0 \nu^{\eta}\ ,
\end{equation}
with $R_0=0.344a$ and $\eta=0.271$ for spheroids and $R_0=0.332a$ and $\eta=0.215$ for the cylinder. We note that the limit of $R$ as $\nu\rightarrow 0$ is not clear from our simulations or on physical grounds. Accessing the linear regime of $\chi(\lambda)$ becomes increasingly difficult as $\nu\rightarrow 0$ and thus $R$ becomes progressively more difficult to determine accurately.

For the cylindrical case, we can compare our results to other proposed expressions. As discussed in the introduction, a commonly used approximation for $R$, from Ref.~\cite{prozorov00}, is
\begin{equation}
	\label{EqRuslanOld}
	R = \frac{a}{2[1+[1+(2\nu)^2]\arctan{((2\nu)^{-1})}-2\nu]} \ ,
\end{equation}
which approaches $R\simeq 0.195a$ in the limit $\nu=0$, and is shown by the dashed red line in Fig.~\ref{figR}(b). The general form of Eq.\ (\ref{EqRuslanOld}) is quite similar to our results, with both rising relatively slowly as $\nu$ is increased. For $\nu$ between 0.05 and 1, the range relevant to typical samples, Eq.\ (\ref{EqRuslanOld}) predicts values of $R$ which are 18\% to 8\% higher than those obtained from our calculations. Therefore experimental works using Eq.\ (\ref{EqRuslanOld}) to extract $\lambda(T)$ from experimental data will have overestimated $\lambda(T)$ by this factor. 

In Fig.\ \ref{figR} we also compare our results for $R(\nu)$ with those obtained recently by Prozorov \cite{prozorov21}. We find that our values of $R(\nu)$ are consistently lower, by up to 10\% despite finding excellent agreement between our respective $\chi(\nu, \lambda)$ data. We have traced the discrepancy to the method of extracting $R$ from $\chi(\lambda)$ used in Ref.\ \cite{prozorov21}. Instead of the quadratic form used here, given by Eq.\ (\ref{EqChi2}), only the first order linear term in $\lambda$ was included in the fit in Ref. \cite{prozorov21}. In principle, if $\chi(\lambda)$ was fitted only over sufficiently small values of $\lambda/a$ this would not introduce a significant error but in the range actually considered in Ref.~\cite{prozorov21} it results in a systematic overestimation of $R$.

For geometries with sharp edges, such as the cylinder considered here, it has been argued that the corner regions may contribute an additional term to the magnetic moment proportional to $\lambda^{2/3}$ \cite{brandt00, mikitik25}. Brandt et al.\ \cite{brandt00} used conformal mapping to find an exact solution for a two-dimensional, perfectly diamagnetic ($\lambda=0$) infinite strip and found that the surface current diverges at the corner. The corresponding finite-$\lambda$ response was then estimated by assuming that this divergence is cut off, and that the current saturates, at a distance of order $\lambda$ from the corner. However, to our knowledge, this assumption has not been tested directly by numerical calculations.

\begin{figure}
	\centering
	\includegraphics[width=\columnwidth]{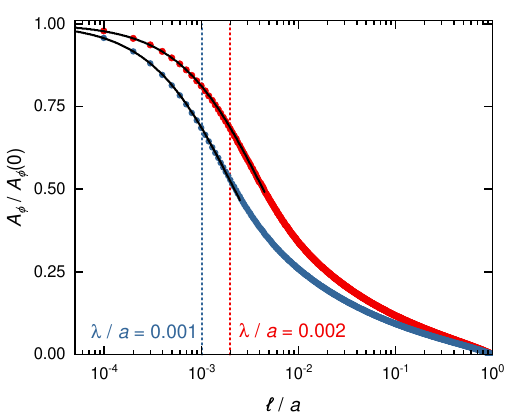}
	\caption{Dependence of the magnetic vector potential, $A_{\phi}$, on the distance from the corner, $\ell$, on the top surface of a cylinder with $\nu=0.05$ and $\lambda/a=0.001$ and $0.002$. The solid line is a fit to $A_{\phi}(\ell)/A_{\phi}(0)=(1+\ell/\ell_c)^{-1}$ up to $2\lambda$, which yields $\ell_c=0.0022$ for $\lambda/a =0.001$ and $\ell_c=0.0043$ for $\lambda/a =0.002$.}
	\label{figAcorner}
\end{figure}

A recent estimate of the magnitude of the supposed $\lambda^{2/3}$ component  \cite{mikitik25} can be compared with the results of our calculations. For the cylinder, this estimate predicts significant curvature in $\chi(\lambda)$ for $\nu<1$. As shown in the inset to Fig.\ \ref{figScaledChi}(b), no such curvature is observed, and the results are well fitted by Eq.\ (\ref{EqChi2}). In Fig.\ \ref{figAcorner}, we plot the radial profile of $A_{\phi}$ on the top surface of a  cylinder with $\nu=0.05$ and $\lambda/a = 0.001$ and 0.002. We find that although $A_{\phi}$ does saturate near the corner, consistent with the qualitative picture suggested in Refs.\ \cite{brandt00, mikitik25},  it approaches this constant value rather slowly and does not  saturate until $\ell \lesssim \lambda/10$ where $\ell$ is the distance from the corner.   The saturation at much smaller distances than those assumed in  Refs.\ \cite{brandt00, mikitik25} is possibly the reason that the   $\lambda^{2/3}$ term determined in those works does not appear in our FEA calculations.  Close to the corner, we find that  $A_{\phi}$ is well-described by $A_{\phi}(\ell)/A_{\phi}(0)\simeq (1+\ell/\ell_c)^{-1}$, and our fit gives $\ell_c\simeq 2\lambda$. This may help refine further calculations similar to those in Refs.\ \cite{brandt00, mikitik25}.

It is possible that for smaller values of $\lambda/a$ than those considered here, some non-linearity of $\chi(\lambda)$ may exist. However, real samples do not have perfectly sharp corners, and rounding on the length scale comparable to or larger than the very small region over which the corner response saturates is almost certain, and would modify this behavior. Hence, even if such a non-linear term emerges at smaller $\lambda/a$, it is likely to be of limited experimental consequence.

\section{Application to Experiments}

For completeness, we outline how the above results can be used to extract changes in the magnetic penetration depth, $\Delta\lambda$, from experiment. Common experimental methods for determining the temperature dependence of $\lambda$ include the tunnel-diode oscillator (TDO) method \cite{giannetta22}, ac susceptibility \cite{bidinosti2000}, SQUID magnetometry \cite{gross1986}, and microwave cavity perturbation \cite{huttema2006}. We focus first on the TDO method, since it provides a direct example of how the geometric factors derived above enter the conversion from a measured signal to $\Delta\lambda$. It is also possible to obtain the absolute value of $\lambda$ using the same techniques \cite{prozorov2000,gross1986,huttema2006}.

In the TDO method, the sample is mounted on a temperature-controlled sapphire rod and inserted into the inductor coil of an LC tank circuit. The circuit is driven by a tunnel diode and typically self-oscillates at a frequency $f_0 \sim 10\,\mathrm{MHz}$. Changes in $\lambda$ alter the magnetic susceptibility of the sample, which changes the effective inductance of the coil and therefore produces a measurable shift in the resonant frequency, $\delta f$.

Assuming that the sample is small compared to the size of the coil, and that the ac-field is uniform over the volume of the sample, $\delta f$ can be related to changes in $\lambda$ by $\delta f = \alpha V_s \delta\chi$, where $\delta\chi$ includes the demagnetization factors and $\alpha\approx f_0/2V_{\rm coil}$ is a calibration constant (where $V_{\rm coil}$ is the volume of the coil). If $\lambda\ll a$, then we can apply the linear form of  Eq.\ (\ref{EqChi2}) and hence
\begin{equation}
\Delta \lambda = -\frac{R (1-N)}{\alpha V_s}\delta f\ .
\end{equation}
This equation can be applied directly, using the above solutions for $N$ and $R$, provided the calibration constant $\alpha$ of the coil is known. $\alpha$ may be determined by measuring $\delta f$ when a small sphere of high conductivity metal, such as aluminium, is placed inside the coil. Here the skin-depth is substituted for $\lambda$ to get the corresponding susceptibility using  Eq.\ (\ref{EqChiSphere}). In practice, however, it is usually preferable to determine the calibration constant in-situ by measuring $\delta f$ when the actual sample being measured is removed from the measurement coil at $T\ll T_c$ where $\lambda \ll a$, which we denote $\Delta f_0$. This accounts both for the demagnetization factor $1/(1-N)$, the sample volume $V_s$ and the coil calibration factor $\alpha$ at the same time, resulting in the simple relation
\begin{equation}
    \Delta \lambda =  R \frac{\delta f}{\Delta f_0}\ .
\end{equation}

The other techniques listed above can be analyzed within the same general framework. Each method relates the measured magnetic response to an effective magnetic volume of the sample, governed by $\chi(\lambda)V_s$ together with the appropriate demagnetizing correction. In ac susceptometry, the measured secondary-coil voltage is proportional to the ac susceptibility of the sample. In SQUID magnetometry, the measured magnetic moment gives the magnetic response directly. In microwave cavity perturbation experiments, the shift of the cavity resonance plays a role analogous to the TDO frequency shift. Thus, although the measured quantity differs between techniques, the conversion from the magnetic response to $\lambda$ or $\Delta\lambda$ is controlled by the same susceptibility, demagnetizing, and effective-dimension factors derived above.

A further consideration is that real experimental samples are rarely ideal cylinders or spheroids. They are more commonly thin plates, parallelepipeds or samples with irregular in-plane geometries. This raises the question of how the measured dimensions should be mapped onto the parameters $c$ and $a$ in the calculations above. Our suggested approach is to use the cylindrical results, taking $2c$ to be the measured sample thickness and taking the effective radius $a=\sqrt{A}/2$, where $A$ is the plate area. The uncertainties in making this approximation can be estimated as follows.

For $B\|c$, in the limit that $c/a\rightarrow \infty$ there are no demagnetization effects and the volume penetrated can be approximated as $\lambda$ times the surface area of the sample perpendicular to the field (as $c/a\rightarrow \infty$ the surfaces parallel to the field can be neglected).  Then, for a cuboid of size $2a\times2b\times2c$, $R_\infty/\tilde{a}=\sqrt{\Gamma}/(1+\Gamma)$, where $\tilde{a}=\sqrt{ab}$, and $\Gamma=b/a$. For $\Gamma=1$,  $R_\infty=0.5\tilde{a}$, the same as for the cylinder. For  $\Gamma=2$ and 4, $R_{\infty}$ increases by 6\%, and 20\% respectively compared to the case with $a=b$. As $R$ varies relatively little with decreasing $\nu$ (Fig.\ \ref{figR}) and as the results for the spheroid and cylinder converge as $\nu$ decreases, we expect the error induced by the $a=\sqrt{A}/2$ approximation to be smaller than the changes in $R_\infty$ as $\Gamma$ increases. Hence for most samples, which typically have $b/a\lesssim 2$, we would expect our suggested approximation to be accurate to $\sim$ 5\%.

Although this approach has not yet been systematically benchmarked against FEA for arbitrary non-axisymmetric geometries, as this would require substantially greater computational resources than were available for the present study, we have performed FEA studies for the high symmetry case of the cube of side $2a$. We find that $R/a=0.3388$, which differs by less than $1\%$ from the value obtained for a cylinder of aspect ratio $c/a=1$. As $R_\infty$ for a cuboid with $a=b$ is also equal to that for a cylinder, this provides further support for the conclusion that $R$ for a cuboid with $a=b$ is unlikely to be significantly different from that of a cylinder as $c/a\rightarrow 0$.

An experimental test of the approximation has been performed in Ref. \cite{carrington2001} using TDO measurements on thin plate-like YBa$_2$Cu$_3$O$_{6+x}$ samples ($a/c \sim 10-20$) with irregular in-plane shape. It was found that using $R=0.2\tilde{a}$, with $\tilde{a}=\sqrt{A}/2$, gave results for $\Delta\lambda(T)$ in the $B\|c$ geometry which differed by $\lesssim$ 5\% from those in the $B\|ab$ geometry. In the latter geometry, there are negligible demagnetizing effects. Although there is a contribution from $c-$axis currents in this orientation this is small because the samples were very thin. For samples of this aspect ratio, our present FEA result for a cylinder differs by only $\sim$2\% from $R=0.2\tilde{a}$. Hence, this suggests that experimentally, $R$ for thin non-cylindrical samples differs from our results for the cylindrical case, with $2a=\sqrt{A}$, by less than 10\%.

A final issue is the nonlinearity of $\chi(\lambda)$ when $\lambda$ is not much smaller than the relevant sample dimension. In most measurements of the temperature dependence of the penetration depth, except very close to $T_c$, the change in penetration depth is small compared with the zero-temperature penetration depth. In that regime, the response can be linearized locally, and $R$ should be interpreted as the inverse local slope relating $\chi$ to $\lambda$ at the relevant value of $\lambda$. Equivalently, the appropriate conversion factor is determined by the derivative of $\chi(\lambda)$ with respect to $\lambda$, evaluated at $\lambda=\lambda(T)$.

As a representative example, consider an optimally doped YBa$_2$Cu$_3$O$_{6+x}$ sample with $T_c=93\,\mathrm{K}$, $\lambda_0\simeq 1000\,\text{\AA}$, and $d\lambda/dT\simeq 5\,\text{\AA}/\mathrm{K}$ \cite{carrington2001}. For typical sample dimensions $2a=0.5\,\mathrm{mm}$ and $2c=25\,\mu\mathrm{m}$, so that $c/a=0.05$, one has $\lambda/a=4\times 10^{-4}$ at $T=0$. At $T=30\,\mathrm{K}$, $\lambda/a$ increases only to approximately $4.6\times 10^{-4}$. Over this range, the fractional difference between the susceptibility change obtained from the linearized form and that obtained by retaining the second-order term is only 0.6\%, corresponding to an error in the inferred $\Delta\lambda$ of approximately $0.9$\,\AA\ over a $150$\,\AA\ change (see Fig.\ \ref{figScaledChi}). However, for samples with much larger $\lambda/a$, or for measurements extending close to $T_c$, nonlinearities in $\chi(\lambda)$ may become important. In those cases, direct inversion of the calculated $\chi(\lambda)$ curves is recommended rather than using the linearized expression. The numerical data for these curves are available \cite{Grant2026Repository}.

\section{Conclusion}

We have presented a finite element study of the spatial variation of the magnetic vector potential of superconducting spheroids and cylinders subject to a uniform magnetic field, explicitly accounting for a finite penetration depth $\lambda$. We find that the magnetic susceptibility is well-described by a second order polynomial in the $\lambda\ll a$ limit, allowing accurate extraction of the effective demagnetization factor, $N$, and the effective dimension, $R$. These results are particularly relevant for the analysis of volume exclusion based measurements of the magnetic penetration depth, where precise knowledge of $N$ and $R$ is essential to convert the results into changes in $\lambda$.

Our analysis shows excellent agreement with analytical and previous numerical results for the demagnetization factor $N$ in both geometries. In contrast we find significant discrepancies with earlier approximate expressions for $R$, and propose a new analytical forms that accurately reproduce our numerical results across the studied parameter range.

\begin{acknowledgments}
We thank Ruslan Prozorov, Joseph Wilcox and Timothy Branch for useful discussions, and RP for sharing data for $\chi(\lambda)$. This work was supported by UK EPSRC grant EP/L015544/1 and carried out in part, using the computational facilities of the Advanced Computing Research Centre, University of Bristol.
\end{acknowledgments}

\bibliography{Demag_2025}

\end{document}